\documentclass[conference]{IEEEtran}

\usepackage{url}

\usepackage[cmex10]{amsmath}

\usepackage{array}

\newtheorem{theorem}{Theorem}

\newtheorem{remark}[theorem]{Remark}

\newcommand{\F}{{\mathbf{F}}} 

\newcommand{\R}{{\mathbf{R}}} 
\newcommand{\PP}{{\mathcal{P}}} 
\newcommand{\C}{{\mathbf{C}}}
\newcommand{\CC}{{\mathcal{C}}}

\newcommand{\SUB}{{\mathcal{S}}}

\newcommand{\Gl}{\operatorname{Gl}}

\newcommand{\Aut}{\operatorname{Aut}}

\newcommand{\Trace}{\operatorname{Trace}}
\newcommand{\wt}{\operatorname{wt}}

\newcommand{\1}{\operatorname{\bf 1}}

\newcommand{\qbinom}[2]{\genfrac{[}{]}{0pt}{}{#1}{#2}}

\newcommand{\rr}{\overline{r}}

\newcommand{\bookhskip}{\hskip 1em plus 0.5em minus 0.4em\relax}

\begin{document}

\title{Applications of Semidefinite Programming to Coding Theory}

\author{
  \IEEEauthorblockN{Christine Bachoc}
  \IEEEauthorblockA{Institut de Math\'ematiques de Bordeaux\\
    Universit\'e Bordeaux 1 \\
    351, cours de la Lib\'eration, 33405 Talence, France\\
    Email: \{christine.bachoc\}@math.u-bordeaux1.fr}
}


\maketitle

\begin{abstract}
  \boldmath
  We survey recent generalizations and improvements of the linear programming
  method that involve semidefinite programming. A general framework
  using group representations and tools from graph theory is provided.
\end{abstract}



\section{Introduction}

The celebrated linear programming method was until recently the most
powerful method to obtain estimates for extremal problems in coding
theory. Initially developed by Philippe Delsarte in the early
seventies in the framework of association schemes \cite{Delsarte1}, it was
proved equally suitable for two-point homogeneous spaces  \cite[Chapter 9]{CS}. We recall that a metric space $X$ with
distance $d_X(x,y)$ is said to be {\em two-point homogeneous} if the group 
of transformations of $X$ preserving $d_X$, which will be called the
automorphism group of $X$ and denoted $\Aut(X)$, acts distance transitively on
$X^2$.
The binary Hamming space $H_n=\{0,1\}^n$ is a  core example of these
spaces, and the fundamental number   $A(n,d)$, the maximal number of
  elements of a binary code of length $n$ with minimal Hamming
  distance at least equal to $d$, can be efficiently upper bounded
  with Delsarte's method. It has lead not only to numerical bounds, 
but also to explicit bounds \cite{Levenshtein},  and to the best known asymptotic
bounds \cite{MRRW}. Moreover this method has been equally  successful
for the other two-point homogeneous spaces 
including
real spaces such as the unit sphere of Euclidean space \cite{DGS}, \cite{KL}. 

Despite of these great results, the search for improvements and for
generalizations of this method  have been fundamental issues
in coding theory. In recent years, results in these directions 
have been obtained using {\em
  semidefinite programming} instead of linear
programming. Semidefinite programming (SDP for short) is a subfield of
convex optimization concerned with the optimization of a linear
functional
over the intersection of the cone of positive semidefinite matrices
with an affine subspace. It contains linear programming (LP) as a
special case. It is a field of recent and growing interest because on
one hand it goes with efficient algorithms, and on the other hand it is
capable of modeling or approximating many optimization problems, in
particular in the area of  combinatorial optimization. 

The first step was taken by A. Schrijver in \cite{Schrijver} who,
using SDP,  was able to improve the known upper bounds for
$A(n,d)$ for some parameters $(n,d)$. In \cite{GST} these results where
extended to the $q$-Hamming space, and in \cite{GMS} they are further
improved. The idea underlying these results is to exploit constraints
involving triples \cite{Schrijver}, \cite{GST}, or even quadruples of codewords
\cite{GMS}.

Generalizing the linear programming method to other metric spaces
$(X,d_X)$ is of special interest in view of the variety of spaces
that play a role in recent areas of coding theory, such as codes over rings,
network coding, and space time coding. The LP method was successfully
generalized to some of these spaces (e.g. the non binary Johnson space
\cite{TAG}, permutation codes \cite{Tarnanen}, the Grassmann spaces \cite{Bachoc1}, \cite{Roy}, the
ordered codes \cite{MartinStinson}, \cite{Barg}, the unitary codes \cite{CreignouDiet}), essentially because
the underlying spaces are {\em symmetric spaces}. Here we mean that 
the group of automorphisms exchanges any two elements of the space. 
However, other spaces do not fit into this framework, such as the
projective space over a finite field or a ball in the Hamming space.

One can go from $H_n$ to another metric space $(X,d_X)$ and ask for
estimates for the analogous number $A(X,d)$. One can also stick to the
Hamming space but consider other types of constraints, such as
constraints involving $k$-tuples of elements instead of pairs. To be
more precise, we want to consider on the
Hamming space $H_n$, some functions $f(x_1,\dots,x_k)$ taking non
negative values, defined on $k$-tuples, that
generalize the Hamming distance. Such a function is called a  {\em
  pseudo-distance}
if it satisfies two properties: is is invariant by a permutation of
the $x_i$ and it is invariant by the diagonal action of $\Aut(H_n)$. Several such pseudo-distances have been
studied in coding theory, for example  the {\em generalized Hamming
  distance} introduced in \cite{CLZ}. Then, one can ask for
$A_{k-1}(n,f,m)$, the maximal
number of elements of a binary code $C$ such that $f(x_1,\dots,x_k)$
is at least equal to some value $m$ when $(x_1,\dots,x_k)$ runs over
the set of $k$-tuples of pairwise distinct elements of $C$.
It can be noticed that $A_1(n,d,m)=A(n,m)$. We show in \cite{BachocZemor} that
Schrijver's method \cite{Schrijver} can be used to derive upper bounds for
$A_2(n,f,m)$ and in \cite{BachocRiener} that it can be generalized to $k\geq 3$.

Our aim in this paper is to give a  general framework for the problems
discussed above, based on a combination of tools from
graph theory,  and from the theory of group representations. 
Indeed, these situations can be interpreted in terms of the
{\em independence number} of specific graphs (or hypergraphs); on the other
hand, an upper bound for the independence number of a graph, which is
the optimal value of a SDP, was discovered by L. Lov\'asz \cite{Lovasz} who
called it the {\em theta number} of the graph. In order to exploit
this upper bound in coding theory, it is necessary to exploit the
action of the automorphism group of the underlying space, and this
step requires tools from 
group representations. We shall illustrate these ideas with the cases
of projective space, Hamming balls and generalized Hamming distance.
In this paper we restrict ourselves to finite spaces, although the
ideas and results extend almost straightforwardly to the case of
compact spaces (see \cite{BGSV}, \cite{BachocVallentin1}).

The paper is organized as follows: Section II reviews Delsarte's LP
method for binary codes. Section III discusses Lov\'asz theta number of
a graph. Section IV introduces semidefinite programs and their
symmetrization.
Section V links the theta number and the LP bound in the case of
binary codes, and introduces the notions relative to a general space $X$.
Section VI gives the necessary results from group representations. Section
VII develops applications to codes in Hamming balls and to codes in
the projective spaces. Section VIII discusses the applications to
pseudo-distances.

\section{Delsarte's LP method for binary codes}
We take the following notations: for $x$ an element of the binary Hamming
space $H_n:=\{0,1\}^n$, the {\em Hamming weight} $\wt(x)$ is the number of
its non zero coordinates and the {\em Hamming distance} of a pair $(x,y)\in
H_n^2$ equals $d_H(x,y)=wt(x-y)$. A certain family of orthogonal
polynomials, the  {\em Krawtchouk polynomials} $K_k^n(t)$,
are 
naturally attached to the Hamming space, and satisfy the so-called
{\em positivity property}:
\begin{equation}\label{pos}
\text{For all } C\subset H_n, \sum_{(x,y)\in H_n^2} K_k^n(d_H(x,y))\geq
0.
\end{equation} 
This property lies at the root of Delsarte LP method. Let us introduce the 
{\em distance distribution} $(x_i)_{0\leq i\leq n}$ of a code $C$:
\begin{equation*}
x_i:=\frac{1}{|C|}|\{(x,y)\in C^2 \ :\ d_H(x,y)=i\}|.
\end{equation*}
Then, these numbers satisfy the following inequalities:
\begin{enumerate}
\item For all $0\leq k\leq n$, $\sum_{i=0}^n K_k^n(i) x_i \geq 0$.
\item $x_i\geq 0$
\item $x_0=1$
\item $\sum_{i=0}^n x_i=|C|$
\end{enumerate}
where 1) rephrases \eqref{pos}. 
Moreover, if $d_H(C)\geq d$, then  $x_i=0$ for $i=1,\dots,d-1$.
From these inequalities, one obtains a linear program in real
variables $y_i$, the optimal
value of which upper bounds the number $A(n,d)$ \cite{Delsarte1}:
\begin{equation}\label{Delsarte LP}
\begin{array}{lll} 
 \max\big\{  \sum_{i=0}^n y_i : &
  y_i\geq 0, \\
&   y_0=1,\\
&   y_i=0 \ \text{ if } i=1,\dots,d-1\\
&  \sum_{i=0}^n K_k^n(i) y_i \geq 0 \quad 0\leq k\leq n \big\}
\end{array}
\end{equation}

In view of generalizations of this method, the role of the
Krawtchouk polynomials should be clarified. In fact, these polynomials 
come into play because they are closely related to the irreducible
decomposition of the space $\CC(H_n)$ of complex valued functions on
$H_n$:
\begin{equation*}
\CC(H_n):=\{f: H_n\to \C\}
\end{equation*}
under the action of the group $\Aut(H_n)$ of transformations of $H_n$
preserving the Hamming distance. This group, of order $2^n n!$,
combines swaps of $0$ and $1$ with permutations of
the coordinates.

More precisely, if $\chi_z(x):=(-1)^{x\cdot z}$ denote the characters
of $(\F_2^n,+)$,
we have:
\begin{equation*}
\CC(H_n)=\oplus_{z\in H_n} \C \chi_z=\oplus_{k=0}^n P_k 
\end{equation*}
where $P_k:=\oplus_{\wt(z)=k} \C\chi_z$ are $\Aut(H_n)$-irreducible subspaces, and the Krawtchouk
polynomials can be  defined by 
\begin{equation*}
K_k^n(d_H(x,y)):=\sum_{\wt(z)=k } \chi_z(x)\chi_z(y)
\end{equation*}
which in turn lead to the explicit expression:
\begin{equation*}
K_k^n(t)=\sum_{j=0}^k (-1)^j \binom{t}{j}\binom{n-t}{k-j}.
\end{equation*}

\section{Lov\'asz's theta number}

Let $\Gamma=(V,E)$ be a finite graph. An {\em independent set} $S$ of $\Gamma$
is a subset of $V$ such that no pair of vertices in $S$ is connected
by an edge, in other words $S^2\cap E=\emptyset$. The {\em
  independence number}  $\alpha(\Gamma)$  is then the maximal number of
elements of an independent set. 

The number $A(n,d)$ studied in coding theory can be interpreted as the
independence number of a particular graph, i.e. the graph $\Gamma(n,d)$
with
vertex set $V=H_n$ and edge set $E=\{(x,y)\in H_n^2 : 0<d_H(x,y)< d\}$.
So the methods developed in graph theory in order to estimate the
independence number can be applied. It turns out that the
exact determination of this graph invariant is a hard problem, but that a
relaxation was defined by L. Lov\'asz \cite{Lovasz} under the name of
the {\em theta number}, which is computable with polynomial complexity
in the size of the graph. More precisely, Lov\'asz theta number
$\vartheta(\Gamma)$ is the optimal value of a {\em semidefinite program}:
\begin{equation*}\label{theta}
\begin{array}{ll} 
  \vartheta(\Gamma) = \max\big\{  \sum_{i,j} B_{i,j} : &
  B=(B_{i,j})_{1\leq i,j\leq v},\ B \succeq 0 \\
&   \sum_i B_{i,i}=1,\\
&   B_{i,j}=0 \quad  (i,j)\in E\big\}
\end{array}
\end{equation*}
where the matrix $B$ is indexed by the vertex set $V=\{1,\dots,v\}$
and where $B\succeq 0$ stands for: $B$ is a symmetric, positive
semidefinite matrix.
The celebrated {\em Sandwich theorem} is proved in \cite{Lovasz}:
\begin{theorem}\label{ST} If $\chi(\overline{\Gamma})$ denotes the chromatic number
  of the complementary graph $\overline{\Gamma}$, then
\begin{equation*}
\alpha(\Gamma)\leq \vartheta(\Gamma)\leq \chi(\overline{\Gamma}).
\end{equation*}
\end{theorem}
\begin{IEEEproof} We only prove the first inequality. Let $S$ be an independent
  set and let $\1_S$ denote its characteristic function. The matrix 
\begin{equation*}
B_{i,j}:=\frac{1}{|S|}\1_S(i)\1_S(j)
\end{equation*}
is feasible for the program $\vartheta(\Gamma)$ and moreover its
optimal value $\sum_{i,j} B_{i,j}$ is equal to $|S|$. So $|S|\leq
\vartheta(\Gamma)$. 
\end{IEEEproof}
However, a straightforward application of the inequality 
$\alpha(\Gamma)\leq \vartheta(\Gamma)$ in the case of binary codes would not be satisfactory 
because the graph $\Gamma(n,d)$ has $2^n$ vertices, thus its size grows
exponentially with the dimension $n$. The key to reduce the
complexity of the computation of $\vartheta(\Gamma(n,d))$ is to exploit the
action of $\Aut(H_n)$ on this graph. In this process, 
up to a minor modification
(i.e. one should consider $\vartheta'(\Gamma(n,d))$, in which the
constraint that 
$B$ takes non negative values is added), 
$\vartheta'(\Gamma(n,d))$ turn to be equal to Delsarte linear programming bound
(see Section V and \cite{Schrijver2}).

The situation described above is in fact very general. In coding
theory, the spaces of interest are always huge spaces, but also
have a huge group of automorphisms. Thus symmetry reduction will play a
crucial role in the design of upper bounds of $\vartheta$ type for extremal problems.

\section{Semidefinite programs}

A semidefinite program (SDP for short) is an optimization problem of
the form:
\begin{equation*}
\begin{array}{ll}
\gamma := \min\big\{ & c_1x_1+\dots +c_mx_m : \\
&  -A_0+x_1A_1+\dots +x_mA_m \succeq 0
\big\}
\end{array} 
\end{equation*}
where $(c_1,\dots,c_m)\in \R^m$,   $A_0,\dots,A_m$ are real  symmetric
matrices,
and the minimum is taken over $(x_1,\dots,x_m)\in \R^m$. 
Linear programs correspond to the special case  of  $A_i$
being diagonal matrices.
The above
{\em primal program} has an associated {\em dual program}, defined
below, where 
$\langle A, B\rangle=\Trace(AB^*)$ is the standard inner product of matrices:
\begin{equation*}
\begin{array}{ll}
\gamma^* := \max\big\{  \   \langle A_0, Z\rangle : & Z\succeq  0, \\
&    \langle A_i, Z\rangle=c_i, \quad i=1,\dots,m
\big\}.
\end{array} 
\end{equation*}
{\em Weak duality}, i.e. $\gamma^*\leq \gamma$, always holds. Under
some mild conditions, one has also {\em strong duality},
i.e. $\gamma=\gamma^*$. In this case, {\em interior point methods} lead to algorithms
that allow to approximate $\gamma$ to an arbitrary precision in
polynomial time. Moreover free solvers are available, e.g. on the
web site NEOS \cite{NEOS}.

Let $G$ be a group of permutations of $\{1,\dots,r\}$. It acts on
matrices of size $r$ by: $(\sigma A)_{i,j}=A_{\sigma^{-1}(i),
  \sigma^{-1}(j)}$, $\sigma\in G$.
The SDP
$\gamma^*$ is said to be {\em $G$-invariant } if the matrices $A_i$ are of
size $r$, if the set $\{Z : Z\succeq  0,  \langle A_i,Z\rangle =c_i\}$
of feasible solutions is globally invariant by $G$, and if
$\sigma A_0=A_0$ for all $\sigma\in G$. In this case, if $Z$ is
a feasible solution, then another feasible solution $Z'$ with the same
optimal value and which is moreover invariant by $G$ is
obtained  by an average of $Z$ on $G$, i.e. setting
\begin{equation*}
Z':=\frac{1}{|G|}\sum_{\sigma\in G} \sigma Z.
\end{equation*}
This reasoning shows that, if $\gamma^*$ is $G$-invariant, one can
restrict the feasible solutions to be $G$-invariant. In other words,
\begin{equation*}
\begin{array}{ll}
\gamma^*=\big(\gamma^*\big)^{G} := \max\big\{ \    \langle A_0,
Z\rangle : 
&  Z\succeq  0, \\
& \sigma Z=Z \text{ for } \sigma\in G,\\
&  \langle A_i, Z\rangle=c_i\ \big\}.
\end{array} 
\end{equation*}
Going from $\gamma^*$ to $\big(\gamma^*\big)^{G}$ is referred to as
{\em symmetry reduction} or {\em symmetrization} of the SDP $\gamma^*$.

\section{Back to the roots}
We come back to $\vartheta'(n,d)$, which is $\Aut(H_n)$-invariant.
Its symmetrization  involves thus 
the matrices $B$ indexed by $H_n$, which are positive semidefinite, and $\Aut(H_n)$-invariant. 
It turns out that there is a beautiful  description of these matrices 
with help of the Krawtchouk polynomials. We now adopt a functional notation
for matrices, i.e. we write $B(x,y)$ instead of $B_{x,y}$.
\begin{theorem}\label{pdf Hamming} $B\in \CC(H_n^2)$ is positive semidefinite and
  $G$-invariant
if and only if
\begin{equation*}
B(x,y)=\sum_{k=0}^d a_k K_k^n(d_H(x,y)) \quad \text{with } a_k\geq 0,
\ 0\leq k\leq d.
\end{equation*}
\end{theorem}
This result shows that the condition $B\succeq 0$ can be replaced by the non
negativity of the variables $a_k$. Replacing in $\vartheta'$, one
obtains a {\em linear program in the variables $(a_0,\dots, a_n)$}.
With a little bit of transformations, one can show that it is equal to
Delsarte linear program.

Now we  consider following the same line for a metric space 
$(X,d_X)$ with group of automorphisms $G$. With the obvious graph
$\Gamma(X,d)$,
we have similarly
\begin{equation}\label{sdp bound} 
A(X,d)\leq \vartheta'(\Gamma(X,d))^{G}.
\end{equation}
Then we need a description 
of the $G$-invariant {\em positive definite functions} $F\in
\CC(X^2)$, i.e such that the matrix $(F(x,y) )_{(x,y)\in X^2}$ is
(Hermitian)
positive semidefinite. This description can be obtained using {\em
  harmonic analysis} of $G$, and is explained in next section.

\section{Tools from harmonic analysis} We shall be rather sketchy
here and refer to \cite{BGSV} for details. In \cite{BGSV},  the more
general case of compact groups is considered. The space $\CC(X)$ is a
$G$-module for the action $(gf)(x):=f(g^{-1}x)$ thus can be decomposed
in irreducible submodules. So we have 
\begin{equation*}
\CC(X)=R_0^{m_0}\perp R_1^{m_1}\perp \dots \perp R_s^{m_s}
\end{equation*}
where the subspaces $R_k$ are pairwise non isomorphic and
$G$-irreducible.
Then, for all $k=0,\dots,s$, one can define a $G$-invariant matrix
$E_k(x,y)$, of size $m_k$, associated to the isotypic subspace
$R_k^{m_k}$, such that we have:
\begin{theorem}
$F\in \CC(X^2)$ is positive definite and $G$-invariant, if and only if 
\begin{equation}\label{psd X}
F(x,y)=\sum_{k=0}^s \langle F_k, E_k(x,y)\rangle \quad \text{ with } F_k\succeq 0.
\end{equation}
\end{theorem}
Moreover, since $E_k(x,y)$ is $G$-invariant, its coefficients only
depend on the orbits $O_G(x,y)$ of pairs $(x,y)\in X^2$ under the
action of $G$, i.e. we have 
\begin{equation*}
E_k(x,y)=Y_k(O_G(x,y))
\end{equation*}
for some matrix $Y_k$. It remains to explicitly compute this matrix,
which is a non trivial task in general. Special cases will be worked
out in the next section. Replacing $F\succeq 0$ by the
expression 
\eqref{psd X} in $\vartheta'(\Gamma(X,d))$ then leads to a semidefinite
program
in the ``variables'' $F_k\succeq 0$. Here we can see exactly when this
SDP turns to be an LP: since the matrices $F_k$ have size $m_k$, it
corresponds to the cases when $m_k=1$ for all $0\leq k\leq s$. One can
show that it is so if $X$ is a {\em symmetric space} as defined in the Introduction.

\section{Bounds for codes in Hamming balls and in Projective geometry}
The projective space $X=\PP_{q,n}$ over $\F_q$, the set of all linear
subspaces of $\F_q^n$, is a metric space for the distance
$d_X(x,y):=\dim(x)+\dim(y)-2\dim(x\cap y)$. Its automorphism group is
the group $G=\Gl_n(\F_q)$ of invertible linear transformations. The
codes of this space have found recent applications in network coding
\cite{Koetter}. Its action on $X$ is not transitive; there are $n+1$  orbits,
the subsets $X_k$ of subspaces of fixed dimension $k$, $0\leq k\leq
n$. The sets $X_k$ themselves are two-point homogeneous, and Delsarte
in \cite{Delsarte2}, who calls them {\em $q$-Johnson spaces}, 
has shown that they can be seen as $q$-analogs of the Johnson spaces,
i.e. the sets of binary words with fixed weight. This analogy in fact
extends to the pairs $(X,G)$ when $X$ is  the full projective space
over $\F_q$
and $G$ is the linear group $\Gl_n(\F_q)$,
respectively the Hamming space and  the symmetric group $S_n$. We
discuss these  situations in a uniform way, with the notations of
\eqref{table}.
\begin{equation}\label{table}
\begin{array}{|c|c|c|}
\hline
X & \PP_{q,n}&  H_n  \\
\hline
q & p^t  & 1\\
G & \Gl_n(\F_q) & S_n \\
|x| & \dim(x) & \wt(x) \\ 
\hline
\end{array}
\end{equation}
$G$ splits $X$ into the orbits $X_k$:
\begin{equation*}
X_k:=\{x\in X : |x|=k\}
\end{equation*}
while the orbits of $X^2$ are:
\begin{equation*}
X_{a,b,c}:=\{ (x,y)\in  X^2 : |x|=a, |y|=b, | x \cap y| =c\}.
\end{equation*}
The distance on these spaces also has a common expression:
\begin{equation*}
d_X(x,y)=|x|+|y|-2|x\cap y|.
\end{equation*}
In \cite{Delsarte2}, the $G$-decomposition of the spaces $\CC(X_k)$ and the
associated polynomials are determined (since the spaces are
two-point homogeneous, the multiplicities $m_k$ are equal to $1$).
They belong to the family of {\em $q$-Hahn polynomials}. From these
results one can go one step further and infer the computation of the
matrices $Y_k$ for the space $X$ (\cite{BachocVallentin2}):
\begin{theorem}\label{BV}
The space $\CC(X)$ contains $1+\lfloor n/2\rfloor$ isotypic subspaces
indexed by $0\leq k\leq \lfloor n/2\rfloor$, with multiplicities
$m_k=n-2k+1$, corresponding to irreducible spaces $R_k$ of dimension $h_k$.
The coefficients of the associated matrices $E_k(x,y)$ are explicitly given
by the formulas:
\begin{equation*} 
E_{k,i,j}(x,y)= |X| h_k\frac{\qbinom{j-k}{i-k}\qbinom{n-2k}{j-k}}{\qbinom{n}{j}\qbinom{j}{i}}q^{k(j-k)}Q_k(n,i,j; i-|x\cap y|)
\end{equation*}
where $k\leq i\leq j\leq n-k$, $|x|=i$, $|y|=j$, 
$E_{k,i,j}(x,y)=0$ if $|x|\neq i$ or $|y|\neq j$, and $Q_k(n,i,j;t)$
are $q$-Hahn polynomials with parameters $n,i,j$.
\end{theorem}
As an application, it is possible to derive from \eqref{sdp bound} upper
bounds for $A(J_{q,n}, d)$ and for $A(B_n(w), d)$ where $B_n(w)$ is
the Hamming ball of radius $w$ centered at $0$:
\begin{equation*}
B_n(w):=\{x\in H_n : \wt(x)\leq w\}.
\end{equation*}
Some numerical results are displayed in Table I. The stars indicate
optimal bounds, attained by the intersection of the Golay code with $B_n(w)$.

\begin{table}[h]\label{Table2}
{\tiny
\begin{equation*}
\begin{array}{|l|l|l|l|l|l|l|l|l|l|}
\hline
n \backslash w  &8&9&10&11&12 & 13 & 14&15&16\\
\hline
18 & 67 & & & & & & &&\\
19 &100 & 123 & 137& & & & && \\
20 & 154& 222 & 253& & & & &&\\ 
21 & 245 & 359 & 465  &   & & & &&\\ 
22 & 349 & 598 & 759 & 870 & 967 & 990 & 1023 &&\\ 
23 & 507 & 831 & 1112 & 1541 & 1800 & 1843 & 1936 & 2047 &\\ 
24 & 760^* & 1161 & 1641& 2419& 3336^* & 3439 & 3711 & 3933 &
4095^* \\ 
\hline
\end{array}
\end{equation*}
}
\caption{SDP bounds for $A(B_n(w), 8)$}
\end{table}

\section{Bounds for binary codes related to pseudo-distances}
We begin with the introduction of three pseudo-distances that have
been studied in coding theory.
For $(x_1,\dots,x_k)\in H_n^k$, the {\em generalized Hamming distance}
$d(x_1,\dots,x_k)$ is defined by:
\begin{equation*}
d(x_1,\dots, x_k)=|\big\{j,\  1\leq j\leq n:\ x^{j} \notin \{0^{k},1^{k}\}\big\}|.
\end{equation*}
where $x^j:= ((x_1)_j,\dots,(x_k)_j)$ denotes the $j$-th column of the
array:
\begin{align*}
x_1&=0\dots01\dots1100\dots0\\
x_2&=0\dots01\dots1011\dots0\\
\vdots \ &  \\
x_k&=0\dots01\dots1\underbrace{001\dots1}_{d(x_1,\dots,x_k)}
\end{align*}
This notion was introduced in \cite{CLZ}, and takes its origin in the
work of Ozarow and
Wyner, and of Wei, who studied the generalized Hamming weight of linear codes
in view of cryptographic applications.
When $k=2$, $d(x_1,x_2)$ is nothing else than the usual Hamming distance.

The {\em radial distance} has connections with the notion of list
decoding (\cite{Bli1}). The radial
distance $r(x_1,\dots,x_k)$ is by definition the smallest radius of a
Hamming ball containing the points $x_1,\dots,x_k$:
\begin{equation*}
r(x_1,\dots,x_k)=\min_{y\in H_n} \big\{\max_{1\leq i\leq k}  d(y,x_i) \big\}.
\end{equation*}
Because this parameter is difficult to analyse, it is sometimes
studied jointly with the {\em average radial distance} (\cite{Bli1})
\begin{equation*}
\rr(x_1,\dots,x_k):=\min_y \big\{\frac{1}{k}\sum_{1\leq i\leq k}  d(y,x_i)\big\}.
\end{equation*}

We want to define an upper bound for the number $A_{k-1}(n,f,m)$
relative to a pseudo-distance $f$, that resembles Lov\'asz's theta
number. In view of the proof of the inequality $\alpha(\Gamma)\leq
\vartheta(\Gamma)$ of Theorem \ref{ST}, it is natural to consider the
function 
\begin{equation}
\chi_C(z_1,\dots,z_k):=\frac{1}{|C|}\1_C(z_1)\dots \1_C(z_k)
\end{equation}
associated to a binary code $C$, and to work out a semidefinite
program from its properties. With this line of thought, we obtain 
in the simplest form:
\begin{theorem}\cite{BachocRiener}\label{BR}
The optimal value of the
following SDP is an upper bound of  $A_{k-1}(n,f,m)$:
\begin{equation*}
\begin{array}{ll}
\max\big\{ \sum_{(x,y)\in H_n^2}F(x,y) :& F:H_n^k\to \R,\\
                                      & F \text{ satisfies }
(1)-(4)\big\}
\end{array}
\end{equation*}
where:
\begin{enumerate}
\item[(1)] $F(z_1,\dots,z_k)=F(\{z_1,\dots,z_k\})$ 
\item[(2)] $(x,y)\mapsto F(x,y,z_3,\dots,z_{k})\succeq 0\text{ and }\geq 0$
\item[(3)] $F(z_1,\dots,z_k)=0$ if $f(z_1,\dots, z_k)\leq m-1$ and
  $z_i\neq z_j$
\item[(4)] $\sum_{x\in H_n} F(x)=1$
\end{enumerate}
\end{theorem}
A slightly stronger condition is used in \cite{BachocRiener} instead
of (2). In order to compute effectively with this program, it is again
necessary
to reduce it with $\Aut(H_n)$, which amounts to express the
$\Aut(H_n)$-invariant functions $F$ satisfying condition (2). This
step can be completed with an analysis  of the positive definite
functions on $H_n$ which are invariant under the stabilizer of $k-2$
elements $(z_3,\dots, z_{k})$. The case $k=3$ corresponds to
the stabilizer of one element, which can be chosen to be the zero
word, thus to the group $S_n$, so this case is contained in Theorem
\ref{BV}. 
The resulting symmetrized program for $k=3$ coincides with the program
used in \cite{Schrijver} (with of course a change in condition
(3)). In \cite{BachocZemor}, numerical bounds for $k=2$ and for the three
pseudo-distances defined above are computed and compared to the
previous known bounds. It turns out that in almost every case the SDP bound
is better. In \cite{BachocRiener}, numerical results are obtained for
$k=4$, i.e. for quadruple functions. However, it seems
difficult to consider larger values of $k$, because the size of the
resulting SDP is of order of magnitude $n^{2^{k-1}-1}$.

\begin{remark} There is also a graphic view point on Theorem
  \ref{BR}. Indeed, the semidefinite program that we have defined,
  upper bounds the independence number of an hypergraph if $H_n$ is
  replaced by its vertex set and if condition (3) is replaced by:
$F(z_1,\dots,z_k)=0$ if $\{z_1,\dots,z_k\}$ is an hyperedge of the
  hypergraph. 
\end{remark}

\begin{remark} The semidefinite bound presented in \cite{GMS}, involves
  functions $F$
defined on the set $\SUB_k(H_n)$ of subsets of $H_n$ of cardinality at
most $k$. The semidefinite constraints on $F$ are as follows: for all $S\subset \SUB_k(H_n)$, 
\begin{equation*}
(X,Y)\mapsto F(X\cup Y)\succeq 0
\end{equation*}
where $X,Y$ run over the elements of $\SUB_k(H_n)$, containing $S$,
and of size at most $(k+|S|)/2$ (so that $|X\cup Y|\leq k$).
The case $|S|=k-2$ corresponds to condition (2). 

The authors obtain with $k=4$ new upper bounds for $A(n,d)$ for sixteen values of
$(n,d)$ in the range $18\leq n\leq 26$ and  $6\leq d\leq 12$. Remarkably, 
the new bound in the case $(n,d)=(20,8)$ reaches the lower bound
provided by successive shortening of the Golay code, thus proves $A(20,8)=256$.
\end{remark}



\enlargethispage{-1cm}




\begin{thebibliography}{3}

\bibitem{Bachoc1}
C. Bachoc,
``Linear programming bounds for codes in Grassmannian spaces'',
{\em IEEE Trans. Inf. Theory}
vol. 52, no. 5 (2006), pp. 2111--2125.

\bibitem{BachocVallentin2}
C. Bachoc and F. Vallentin,
``More semidefinite programming bounds'', 
in {\em Proceeding of DMHF 2007}, Fukuoka, 2007.


\bibitem{BachocVallentin1}
C. Bachoc and F. Vallentin,
``New upper bounds for kissing numbers from semidefinite programming'',
{\em J. Amer. Math. Soc}, vol.  21 (2008), pp. 909--924.


\bibitem{BachocZemor}
C. Bachoc and G. Z\'emor,
``Bounds for binary codes relative to pseudo-distances of $k$
points'',  to appear in Adv. Math. Com.

\bibitem{BachocRiener}
C. Bachoc and C. Riener,  in preparation.




\bibitem{BGSV} C. Bachoc, D. Gijswijt, A. Schrijver and F. Vallentin,
``Invariant semidefinite programs'', in preparation.

\bibitem{Barg}
A. Barg and  P. Purkayastha,
``Bounds on ordered codes and orthogonal arrays'',  
{\em Moscow Math. Journal} vol. 2 (2009).

\bibitem{Bli1} V. M. Blinovskii,
``Bounds for codes in the case of list decoding of finite volume'',
{\em Problems of Information Transmission}, vol. 22, no. 1 (1986),
pp. 7--19.




\bibitem{CLZ} G. Cohen, S. Litsyn and  G. Z\'emor,
``Upper bounds on generalized Hamming distances'', 
{\em IEEE Trans. Inf. Theory}, vol. 40 (1994), 2090-2092.

\bibitem{CS} J.H. Conway and N.J.A. Sloane, 
{\em Sphere Packings, Lattices and Groups}.\bookhskip
Springer-Verlag, 1988.


\bibitem{CreignouDiet} J. Creignou and H. Diet,
``Linear programming bounds for unitary codes'', 
{em Advances in Math. Com.}, to appear in Adv. Math. Com.

\bibitem{Delsarte1} P. Delsarte, 
``An algebraic approach to the association schemes of coding
  theory'',
{\em Philips Res. Rep. Suppl.} (1973), vi+97.

\bibitem{Delsarte2} P. Delsarte, 
``Hahn polynomials, discrete harmonics and $t$-designs'',
{\em SIAM J. Appl. Math.}, vol. 34, no. 1 (1978).


\bibitem{DGS} P. Delsarte, J.M. Goethals and  J.J. Seidel,
``Spherical codes and designs'',
{\em Geom. Dedicata}, vol. 6  (1977), pp. 363--388.



\bibitem{GST} D.C. Gijswijt, A. Schrijver and  H. Tanaka, 
``New upper bounds for nonbinary codes'', 
{\em J. Combin. Th. Ser. A}, vol. 13 (2006), pp. 1717--1731. 

\bibitem{GMS} D.C. Gijswijt, H. Mittelmann and A. Schrijver,
``Semidefinite code bounds based on quadruple distances'',
arXiv:1005.4959


\bibitem{KL} G.A.~Kabatiansky and  V.I.~Levenshtein,
``Bounds for packings on a sphere and in space'',
{\em Problems of Information Transmission}, vol. 14 (1978), pp. 1--17.



\bibitem{Koetter} R. Koetter,
``Coding for errors and erasures in random network coding'',
 in {\em Proc. IEEE Int. Symp. Information Theory}, 2007.


\bibitem{Levenshtein} V. I. Levenshtein,
"Universal bounds for codes and designs", in
{\em Handbook of Coding Theory}, eds V. Pless and W. C. Huffmann,
Amsterdam: Elsevier, 1998, pp. 499--648.

\bibitem{Lovasz} L. Lov\'asz, 
``On the Shannon capacity of a  graph'', 
{\em IEEE Trans. Inform. Theory} vol. 25 (1979), pp. 1-5.



\bibitem{MartinStinson}
W.J. Martin and D.R. Stinson, 
``Association schemes for ordered orthogonal arrays and $(T,M,S)$-nets'',
{\em Canad. J. Math.} vol. 51, no. 2 (1999), pp. 326--346. 


\bibitem{MRRW}  R. J. McEliece, E. R.  Rodemich, H. Rumsey, L.  Welch,
"New upper bounds on the rate of a code via the Delsarte-MacWilliams
inequalities", {\em IEEE Trans. Inf. Theory} vol. 23 (1977),
pp. 157--166.


\bibitem{NEOS} http://www-neos.mcs.anl.gov/

\bibitem{Roy} A. Roy,
``Bounds for codes and designs in complex  subspaces'',
preprint, arXiv:0806.2317

\bibitem{Schrijver2} A. Schrijver, 
``A comparaison of the Delsarte  and Lov\'asz bound'', 
{\em IEEE Trans. Inform. Theory} vol. 25 (1979), pp. 425--429.




\bibitem{Schrijver} A. Schrijver, ``New code upper bounds from
the Terwilliger algebra and semidefinite programming'', 
{\em IEEE Trans. Inf. Theory} vol. 51 (2005), pp. 2859--2866.

\bibitem{TAG} H. Tarnanen, M.  Aaltonen and J.-M.  Goethals,
``On the nonbinary Johnson scheme'', 
{\em  Europ. J. Comb.},  vol. 6, no. 3  (1985), pp. 279--285. 

\bibitem{Tarnanen} H. Tarnanen,
``Upper bounds on permutation codes via linear  programming'',
{\em Europ. J. Comb.}, vol. 20, (1999), pp.  101--114.




\end{thebibliography}


\end{document}